\documentclass[english,superscriptaddress, notitlepage, breaklinks=true, showkeys, showpacs=false, nofootinbib, prl]{revtex4}
\usepackage[T1]{fontenc}
\usepackage[latin9]{inputenc}
\usepackage{color}
\usepackage{babel}
\usepackage{amsmath}
\usepackage{amssymb}
\usepackage{graphicx}
\usepackage[unicode=true,pdfusetitle,
 bookmarks=true,bookmarksnumbered=false,bookmarksopen=false,
 breaklinks=true,pdfborder={0 0 0},backref=false,colorlinks=true]
 {hyperref}
\usepackage{breakurl}

\makeatletter
\@ifundefined{textcolor}{}
{%
 \definecolor{BLACK}{gray}{0}
 \definecolor{WHITE}{gray}{1}
 \definecolor{RED}{rgb}{1,0,0}
 \definecolor{GREEN}{rgb}{0,1,0}
 \definecolor{BLUE}{rgb}{0,0,1}
 \definecolor{CYAN}{cmyk}{1,0,0,0}
 \definecolor{MAGENTA}{cmyk}{0,1,0,0}
 \definecolor{YELLOW}{cmyk}{0,0,1,0}
}

\hypersetup{colorlinks=true,citecolor=blue,linkcolor=cyan,urlcolor=blue,filecolor= green, breaklinks=true}
\allowdisplaybreaks
\usepackage{url}
\usepackage[braket, qm]{qcircuit}
\usepackage{listings}

\makeatother

\begin{document}

\title{Preparing tunable Bell-diagonal states on a quantum computer}

\author{Mauro B. Pozzobom}

\address{Departamento de F\'isica, Centro de Ci\^encias Naturais e Exatas, Universidade Federal de Santa Maria, Avenida Roraima 1000, Santa Maria, RS, 97105-900, Brazil}

\author{Jonas Maziero}

\email{jonas.maziero@ufsm.br}

\address{Departamento de F\'isica, Centro de Ci\^encias Naturais e Exatas, Universidade Federal de Santa Maria, Avenida Roraima 1000, Santa Maria, RS, 97105-900,  Brazil}
\begin{abstract}
The class of two-qubit Bell-diagonal states has been the workhorse
for developing understanding about the geometry, dynamics, and applications
of quantum resources. In this article, we present a quantum circuit
for preparing Bell-diagonal states on a quantum computer in a tunable
way. We implement this quantum circuit using the IBM Q 5 Yorktown
quantum computer and, as an application example, we measure the non-local,
steering, entanglement, and discord quantum correlations and non-local
quantum coherence of Werner states.
\end{abstract}

\keywords{Bell-diagonal states; Werner states; quantum computer; IBM Q 5 Yorktown;
quantum resources}

\pacs{42.50.Dv; 03.65.Ud; 03.65.Yz; 03.67.-a}

\maketitle
Quantum properties such as coherence \cite{plenio}, nonlocality \cite{Bell},
steering \cite{Schrodinger}, entanglement \cite{Werner}, and discord
\cite{Zurek} have been identified as resources enabling the implementation
of diverse quantum computation and communication protocols \cite{Shi,Brunner,Branciard,Popescu,Streltsov}.
The functions defined to quantify these quantum features based on
the resource-theory framework \cite{plenio,brito,gallego,vedral,lloyd}
are frequently hard to compute analytically for general quantum states
\cite{gharibian,huang}. Motivated by that observation, a subset of
two-qubit states, the so called Bell-diagonal states (BDS), have been
used extensively for better understanding some of these resources
\cite{horodecki,lang,quan,roszak,meng,kay,wang,singh,
maziero,celeri,castro,paula,luo,hou,han,bromley,bromley2,
du,auccaise,auccaise2,aguilar}.

So, due to its central place within the study of quantum resources,
the experimental preparation of BDS is of apparent need. Recently
Liu \emph{et al.} showed how to prepare tunable Werner states in a
linear optical system via the implementation of a depolarizing channel
applied to a Bell state \cite{liu}. Here we devise a simple quantum
circuit that can be used to create tunable BDS on a quantum computer
with the use of two auxiliary qubits. To exemplify the use of our
protocol, we measure experimentally, using the IBM Q 5 Yorktown quantum
computer \cite{ibm}, the quantum nonlocality, steering, entanglement,
discord, and non-local coherence of Werner states, which are a one-parameter
subset of the BDS.

Our protocol can find application in verifying experimentally several
theoretical results from the literature. For instance, one can apply
our circuit to verify the relation between the sudden change phenomenon
of quantum discord and the worst case fidelity in quantum teleportation,
discovered in \cite{roszak}. The necessity of quantum entanglement,
instead of quantum non-locality, for better than classical fidelity
of quantum teleportation exemplified using the thermal state associated
with the magnetic dipolar interaction Hamiltonian \cite{castro} can
also be simulated using our protocol. This procedure can also be applied
to verify the direct-dynamical entanglement-discord relations reported
in \cite{feldman}. Besides these three examples, one can easily find
several other applications for our protocol, as e.g. in the experimental
verification of the theoretical results reported in Refs. \cite{kay,li,han,lang,li2}.

The remainder of this article is organized as follows. We begin describing
the class of BDS and presenting the quantum circuit we propose for
its preparation on a quantum computer (QC). In the sequence we outline
the implementation of this circuit on the IBM Q 5 Yorktown QC, hereafter
referred to as ibmqx2. Then we present the experimental results we
obtained for the quantum correlations and non-local coherence of Werner
states. Finally, we report on a simple model that we have introduced
to explain the noise influence on the experimental data and give our
conclusions.

\begin{figure}
\begin{centering}
\centerline{
\Qcircuit @C=1em @R=0.7em
{
\lstick{\ket{0}_{a}} & \gate{R(\theta/2)} & \ctrl{2} & \qw      & \qw      & \qw      & \qw \\
\lstick{\ket{0}_{b}} & \gate{R(\alpha/2)} & \qw      & \ctrl{2} & \qw      & \qw      & \qw \\
\lstick{\ket{0}_{c}} & \qw                    & \targ    & \qw      &  \qw & \targ  & \qw \\
\lstick{\ket{0}_{d}} & \qw                    & \qw      & \targ    & \gate{H}  & \ctrl{-1}  & \qw 
}
}
\par\end{centering}

\caption{Quantum circuit we propose to generate tunable Bell-diagonal states
on a quantum computer. The $R$ gates generate one-qubit superposition
states. The CNOTs are used to copy the states of the qubits $a$ and
$b$ to the qubits $c$ and $d$, respectively. By its turn, the Hadamard
and the last CNOT gate are used for changing from the computational
to the Bell basis. At the output, the joint state of qubits $c$ and
$d$ is equivalent to $\rho^{bd}$ of Eq. (\ref{eq:bds}) with $p_{jk}$
given as in Eq. (\ref{eq:prob}).}

\label{QC_BDS}
\end{figure}
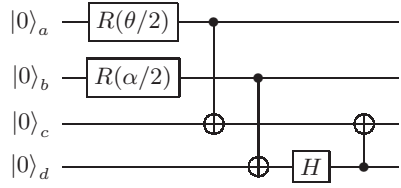

As the name indicates, two-qubit Bell-diagonal states read
\begin{equation}
\rho^{bd}=\sum_{j,k=0}^{1}p_{jk}|\beta_{jk}\rangle\langle\beta_{jk}|,\label{eq:bds}
\end{equation}
where $|\beta_{jk}\rangle=2^{-1/2}(|0\rangle\otimes|k\rangle+(-1)^{j}|1\rangle\otimes|k\oplus1\rangle)$
are the Bell's base states \cite{nilsen}, with $\oplus$ being
the modulo $2$ sum, $\{|j\rangle\}_{j=0}^{1}$ is the computational
basis, and $p_{jk}$ is a probability distribution. This class of
two-qubit states has the following four-qubit purification
\begin{equation}
|\tau\rangle_{abcd}=\sum_{j,k=0}^{1}\sqrt{p_{jk}}|j\rangle_{a}\otimes|k\rangle_{b}\otimes|\beta_{jk}\rangle_{cd}.
\end{equation}
That is to say,
\begin{equation}
\rho^{bd}=\rho_{cd}=\mathrm{Tr}_{ab}(|\tau\rangle\langle\tau|_{abcd}),
\end{equation}
with $\mathrm{Tr}_{ab}$ being the partial trace function \cite{mazieroPTr}.
Here we report that the quantum circuit shown in Fig. \ref{QC_BDS}
generates the $4$-qubit state $|\tau\rangle_{abcd}$, and therefore
that it can be used to prepare any BDS.

In the circuit shown in Fig. \ref{QC_BDS}, we used the rotation
\begin{equation}
R(x)=\begin{bmatrix}\cos x & -\sin x\\
\sin x & \cos x
\end{bmatrix},
\end{equation}
the controlled-not gate
\begin{equation}
CNOT_{s\rightarrow s'}=|0\rangle\langle0|_{s}\otimes\sigma_{0}^{s'}+|1\rangle\langle1|_{s}\otimes\sigma_{1}^{s'},
\end{equation}
and the Hadamard gate
\begin{equation}
H=\frac{1}{\sqrt{2}}\begin{bmatrix}1 & 1\\
1 & -1
\end{bmatrix}.
\end{equation}
Above $\sigma_{0}^{s}\equiv\sigma_{0}$ is the $2\mathrm{x}2$ identity
matrix and $\{\sigma_{j}^{s}\equiv\sigma_{j}\}_{j=1}^{3}$ are the
Pauli matrices acting on qubit $s$ \cite{nilsen}.

For the circuit in Fig. \ref{QC_BDS}, the relations among the probabilities
in the BDS and rotation angles are seen to be
\begin{equation}
p_{jk}=\left(\cos^{2}\frac{\theta}{2}\right)^{1-j}\left(\sin^{2}\frac{\theta}{2}\right)^{j}\left(\cos^{2}\frac{\alpha}{2}\right)^{1-k}\left(\sin^{2}\frac{\alpha}{2}\right)^{k}.\label{eq:prob}
\end{equation}

For the calculation of quantum correlations, one usually start studying
a maximally-mixed marginals state, $\rho_{3m}=2^{-2}(\sigma_{0}\otimes\sigma_{0}+\sum_{j,k=1}^{3}c_{jk}'\sigma_{j}\otimes\sigma_{k})$,
put to the normal form \cite{luo2},
\begin{equation}
\rho_{n}=2^{-2}\left(\sigma_{0}\otimes\sigma_{0}+\sum_{j=1}^{3}c_{jj}\sigma_{j}\otimes\sigma_{j}\right),
\end{equation}
via local unitaries. The states $\rho_{n}$ are diagonal in the Bell
basis, having the following eigenvalue--eigenvector pairs 
\begin{equation}
\left(p_{jk}=\frac{1}{4}\left(1+(-1)^{j}c_{1}+(-1)^{j+k-1}c_{2}+(-1)^{k}c_{3}\right)\mbox{, }|\beta_{jk}\rangle\right),\label{eq:prang}
\end{equation}
where we used $c_{j}\equiv c_{jj}$. 

Hence, from Eqs. (\ref{eq:prob}) and (\ref{eq:prang}) we see that
given $\rho_{n}$, we can prepare any BDS in a tunable way by using
as input to the quantum computer rotations $R(\theta/2)$ and $R(\alpha/2)$
the angles:
\begin{eqnarray}
\theta & = & 2\arccos\sqrt{p_{00}+p_{01}}\mbox{,}\\
\alpha & = & 2\arccos\sqrt{p_{00}+p_{10}}.
\end{eqnarray}

For the implementation of the quantum circuit of Fig. \ref{QC_BDS}
on the ibmqx2, we use $R(x/2)=U_{3}(x,0,0)$ with

\begin{equation}
U_{3}(\theta,\phi,\lambda)=\begin{bmatrix}\cos\frac{\theta}{2} & -e^{i\lambda}\sin\frac{\theta}{2}\\
e^{i\phi}\sin\frac{\theta}{2} & e^{i(\lambda+\phi)}\cos\frac{\theta}{2}
\end{bmatrix}
\end{equation}
being one of the ibmqx2 quantum gates \cite{ibm}. The other gates
we need are themselves directly included in the ibmqx2 set of ready-to-use
quantum gates. 

The experiments were carried out with the calibration parameters for
the ibmqx2 shown in Table \ref{calibration}. We have chosen the following
encoding (see Table \ref{calibration}) for implementation of the
circuit in Fig. \ref{QC_BDS}:
\begin{equation}
a\rightarrow Q1\mbox{, }b\rightarrow Q3\mbox{, }c\rightarrow Q2\mbox{, }d\rightarrow Q4.
\end{equation}

\begin{table} 
\begin{tabular}{|c|c|c|c|c|c|}
\hline ibmqx2 parameters averages  & Q0 & Q1 & Q2 & Q3 & Q4\tabularnewline \hline  \hline  Frequency (GHz) & 5.29 & 5.23 & 5.02 & 5.29 & 5.08\tabularnewline \hline  T1 ($\mu$s) & 50.81 & 59.80 & 64.93 & 56.37 & 56.81\tabularnewline \hline  T2 ($\mu$s) & 45.89 & 39.70 & 63.14 & 31.60 & 32.32\tabularnewline \hline  Gate error ($10^{-3}$) & 2.82 & 1.83 & 4.65 & 4.36 & 2.54\tabularnewline \hline  Readout error ($10^{-2}$) & 4.16 & 1.89 & 1.93 & 2.87 & 4.61\tabularnewline \hline  MultiQubit gate error ($10^{-2}$) & $\stackrel{CX0\_1}{4.15}$ & $\stackrel{CX1\_2}{3.81}$ &  & $\stackrel{CX3\_2}{7.09}$ & $\stackrel{CX4\_2}{3.84}$\tabularnewline \hline   & $\stackrel{CX0\_2}{4.42}$ &  &  & $\stackrel{CX3\_4}{5.28}$ & \tabularnewline \hline 
\end{tabular}
\caption{Averages of the calibration data of the IBM Q 5 Yorktown quantum computer with which the experiments were performed. The temperature was $T=0.0159\text{ K}$.}
\label{calibration}
\end{table}

With these settings, we prepared Werner states \cite{Werner},
\begin{equation}
\rho_{w}=(1-w)\frac{\sigma_{0}\otimes\sigma_{0}}{4}+w|\beta_{11}\rangle\langle\beta_{11}|,
\end{equation}
for eleven values of $w\in[0,1]$. We observe that $\rho_{w}$ is
equivalent to $\rho_{n}$ if $c_{1}=c_{2}=c_{3}=-w$. 

\begin{figure}
\includegraphics[scale=0.6]{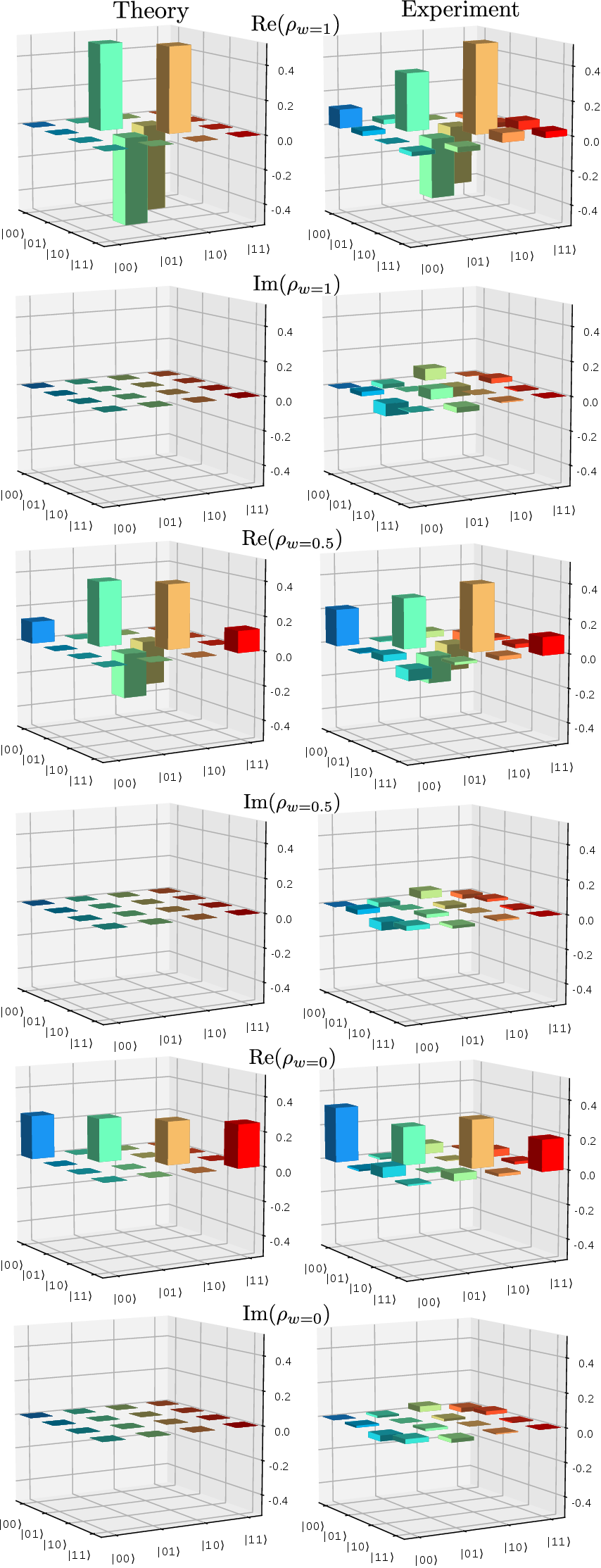}

\caption{(color online) Computational basis representation of the real and
imaginary parts of the theoretical-target and experimentally prepared
density matrices corresponding to Werner states with three different
values of the weight $w$.}

\label{tomo}
\end{figure}

In order to experimentally reconstruct the prepared states, we consider
general two-qubit states written in the form
\begin{equation}
\rho=\frac{1}{4}\sum_{j,k=0}^{3}c_{jk}\sigma_{j}\otimes\sigma_{k},
\end{equation}
with $c_{jk}=\langle\sigma_{j}\otimes\sigma_{k}\rangle_{\rho}$. All
of these averages can be obtained from the joint probability distributions
of the local measurements of $\sigma_{j}$ and $\sigma_{k}$. Let
\begin{equation}
p_{j\pm,k\pm}:=\mathrm{Prob}(\sigma_{j}=\pm1,\sigma_{k}=\pm1).
\end{equation}
Then, for $j,k=1,2,3$:
\begin{equation}
c_{jk}=p_{j+,k+}+p_{j-,k-}-p_{j+,k-}-p_{j-,k+}.
\end{equation}
Using the marginal probability distributions
\begin{equation}
p_{j\pm}=p_{j\pm,k+}+p_{j\pm,k-}\mbox{ and }p_{k\pm}=p_{j+,k\pm}+p_{j-,k\pm}
\end{equation}
we calculate
\begin{equation}
c_{j0}=p_{j+}-p_{j-}\mbox{ and }c_{0k}=p_{k+}-p_{k-}
\end{equation}
for $j=1,2,3$ and for $k=1,2,3$. Finally, because $\mathrm{Tr}(\rho)=1$,
we have $c_{00}=1$. Measurements of $\sigma_{3}$ are part of the
ready-to-use operations of ibmqx2. To measure $\sigma_{1}$, we first
applied the Hadamard gate $H$ and then measured $\sigma_{3}$. For
measuring $\sigma_{2}$, we applied
\begin{equation}
S^{\dagger}=\begin{bmatrix}1 & 0\\
0 & -i
\end{bmatrix},
\end{equation}
then applied $H$, and finally measured $\sigma_{3}$. Above $\dagger$
denotes the transpose conjugate. With these measurement procedures,
the probability distributions $p_{j\pm,k\pm}$ were estimated with
$8192$ runs of the given quantum circuit and corresponding measurements.
The computational basis representation of the reconstructed Werner
states density matrices is shown in Fig. \ref{tomo} for three values
of $w$.

In the sequence, we shall describe the quantumness measures we consider
in this article. We begin by the $l_{1}$-norm coherence, with the
standard basis used as the reference basis \cite{plenio}:
\begin{equation}
C_{l_{1}}(\rho)=\sum_{j\ne k}|\langle j|\rho|k\rangle|
\end{equation}
for $j,k=1,\cdots,d$ with $d$ being the dimension of the regarded
state space. A natural candidate for quantifying the non-local extent
of the quantum coherence of a bipartite system is \cite{pozzobom,byrnes}:
\begin{equation}
C(\rho)=C_{l_{1}}(\rho)-[C_{l_{1}}(\rho_{a})+C_{l_{1}}(\rho_{b})],
\end{equation}
with the reduced states given by \cite{mazieroPTr}: $\rho_{a}=\mathrm{Tr}_{b}(\rho)$
and $\rho_{b}=\mathrm{Tr}_{a}(\rho)$. 

The quantum correlation (QC) named quantum discord is related to the
minimal extent to which the correlations in a composite system are
to be deprecated by local non-selective projective measurements. Here
we use Ollivier-Zurek's discord \cite{Zurek}:
\begin{equation}
D(\rho)=I(\rho)-\max_{\Pi_{b}}I(\Pi_{b}(\rho)),
\end{equation}
with the quantum mutual information being $I(x)=S(x_{a})+S(x_{b})-S(x)$,
where $x_{a}$ and $x_{b}$ are reduced operators computed as mentioned
above. By its turn, the measured state is defined as $\Pi_{b}(\rho)=\sum_{j}\sigma_{0}\otimes\Pi_{j}^{b}\rho\sigma_{0}\otimes\Pi_{j}^{b}$
with $\Pi_{j}^{b}\Pi_{k}^{b}=\delta_{j,k}\Pi_{j}^{b}$ and $\sum_{j}\Pi_{j}^{b}=\sigma_{0}$.
We observe that once there is no known analytical formula for $D$
of general states (even for two qubits), the results we present in
the sequence are obtained using numerical optimization.

Discord is known to be a weaker quantum correlation when compared
to entanglement. This last type of quantum correlation, the non-separable
correlations, are quantified here using the entanglement negativity
\cite{vidal}:
\begin{equation}
E(\rho)=||T_{p}(\rho)||_{tr}-1,
\end{equation}
where $||X||_{tr}=\mathrm{Tr}\sqrt{X^{\dagger}X}$ is the trace norm
and $T_{p}$ is the partial transposition operation \cite{mazieroPT}.

For the two strongest forms of quantum correlations known, steering
and non-locality, which cannot be explained using a local hidden state
and a local hidden variable model, respectively, we use the formulas
reported in \cite{costa}. These authors considered measures for these
quantities given by the maximum extend to which a given related inequality
\cite{chsh,horodecki2,ecavalcanti} is violated. For deriving their
analytical formulas, they used the standard form for two-qubit states
\cite{luo2}:
\begin{equation}
4\rho=\sigma_{0}\otimes\sigma_{0}+\vec{a}\cdot\vec{\sigma}\otimes\sigma_{0}+\sigma_{0}\otimes\vec{b}\cdot\vec{\sigma}+\sum_{j=1}^{3}c_{j}\sigma_{j}\otimes\sigma_{j}.
\end{equation}
This form can be obtained via local unitary transformations applied
locally to a general two-qubit state, i.e., $\rho=U_{a}\otimes U_{b}\rho_{g}U_{a}^{\dagger}\otimes U_{b}^{\dagger}$,
for $U_{a}U_{a}^{\dagger}=U_{a}^{\dagger}U_{a}=U_{b}U_{b}^{\dagger}=U_{b}^{\dagger}U_{b}=\mathbb{I}$,
with
\begin{equation}
4\rho_{g}=\sigma_{0}\otimes\sigma_{0}+\vec{x}\cdot\vec{\sigma}\otimes\sigma_{0}+\sigma_{0}\otimes\vec{y}\cdot\vec{\sigma}+\sum_{j,k=1}^{3}c_{jk}\sigma_{j}\otimes\sigma_{k},
\end{equation}
 where $\vec{a}=O_{a}\vec{x}$, $\vec{b}=O_{b}\vec{y}$, and $\mathrm{diag}(c_{1},c_{2},c_{3})=O_{a}CO_{b}^{T}$,
with $C=(c_{jk})$ being the correlation matrix and $O_{a}$ and $O_{b}$
are orthonormal matrices, i.e., $O_{a}O_{a}^{T}=O_{a}^{T}O_{a}=O_{b}O_{b}^{T}=O_{b}^{T}O_{b}=\mathbb{I}$
for $x^{T}$ denoting the transpose of the matrix $x$. The authors
of \cite{costa} obtained analytically the steering for three measurements
per qubit, 
\begin{equation}
S(\rho)=\max\left(0,\frac{||\vec{c}||-1}{\sqrt{3}-1}\right),
\end{equation}
and the quantum non-locality for two measurements per qubit,
\begin{equation}
N(\rho)=\max\left(0,\frac{\sqrt{||\vec{c}||^{2}-c_{\min}^{2}}-1}{\sqrt{2}-1}\right),
\end{equation}
with $c_{\min}$ being the minimum value among the components of the
correlation vector $\vec{c}=(c_{1},c_{2},c_{3})$. Here we use as
the correlation vector the singular values of the correlation matrix
$C=(c_{jk})$, for $j,k=1,2,3$. We emphasize that the standard form
is obtained via local unitary transformations, which do not affect
the non-locality and steering functions above. Besides, we utilize
the original state (reconstructed or theoretical) for the calculation
of non-local coherence, discord, and entanglement.

\begin{figure}
\includegraphics[scale=0.8]{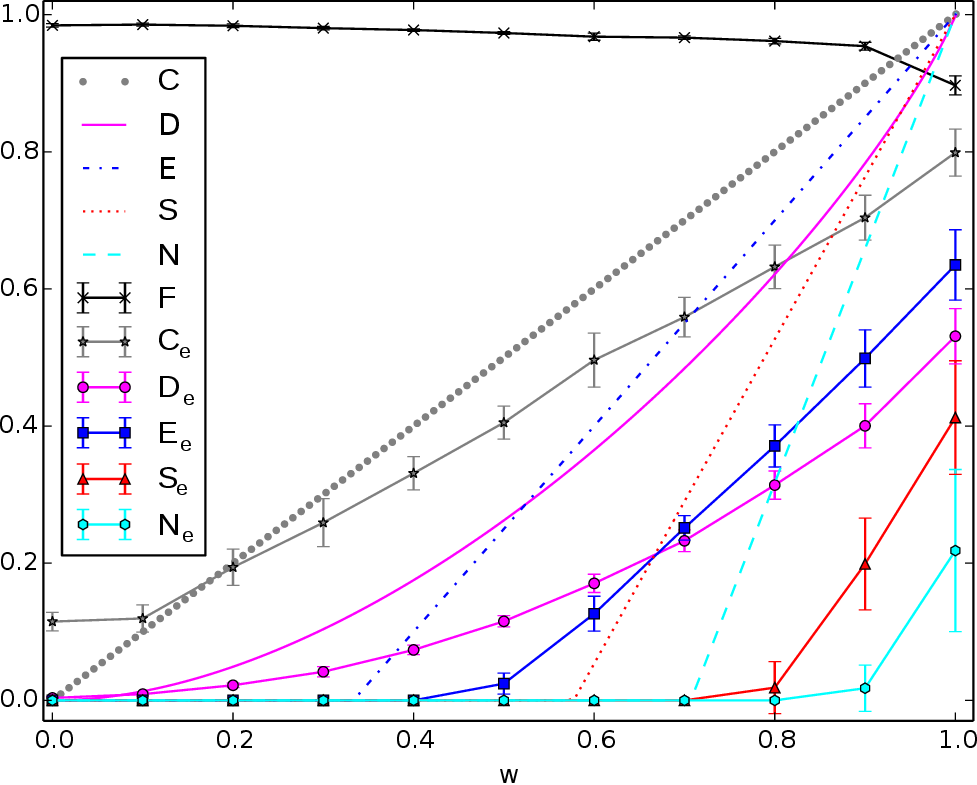}

\caption{(color online) The \textbf{x}-like black points show the preparation
fidelities, $F(\rho_{w},\rho_{w}^{exp})=Tr\sqrt{\sqrt{\rho_{w}}\rho_{w}^{exp}\sqrt{\rho_{w}}}$.
The fidelity and the functions indicated in the legend by the subscript
$e$ refer to averages, and the associated standard deviations, computed
for the experimentally prepared states using seven rounds of experiments.
$C$ stands for non-local coherence, shown in gray. In magenta is
plotted the Ollivier-Zurek discord $D$. The entanglement negativity
$E$ is shown in blue. The steering $S$ was given the color red and
the non-locality $N$ is shown in cyan.}

\label{qcorr}
\end{figure}

The results for the state preparation fidelity and for all these quantum
non-local coherence and correlation measures are presented in Fig.
\ref{qcorr}. The code we used to compute these functions is freely
available at \href{https://github.com/jonasmaziero/libPyQ}{https://github.com/jonasmaziero/libPyQ}.

Even though the preparation fidelities shown in Fig. \ref{qcorr}
have, in general, values quite close to the maximum value $1$, we
see in this figure that the environmental noise and the quantum computer
imperfections have significant detrimental effects on the quantum
properties of the prepared states. This fact indicates that state
preparation fidelity is not a reliable figure of merit if ones main
purpose is the production and utilization of quantum correlations.
For a related discussion, see Ref. \cite{mandarino}.

\begin{figure}
\includegraphics[scale=0.77]{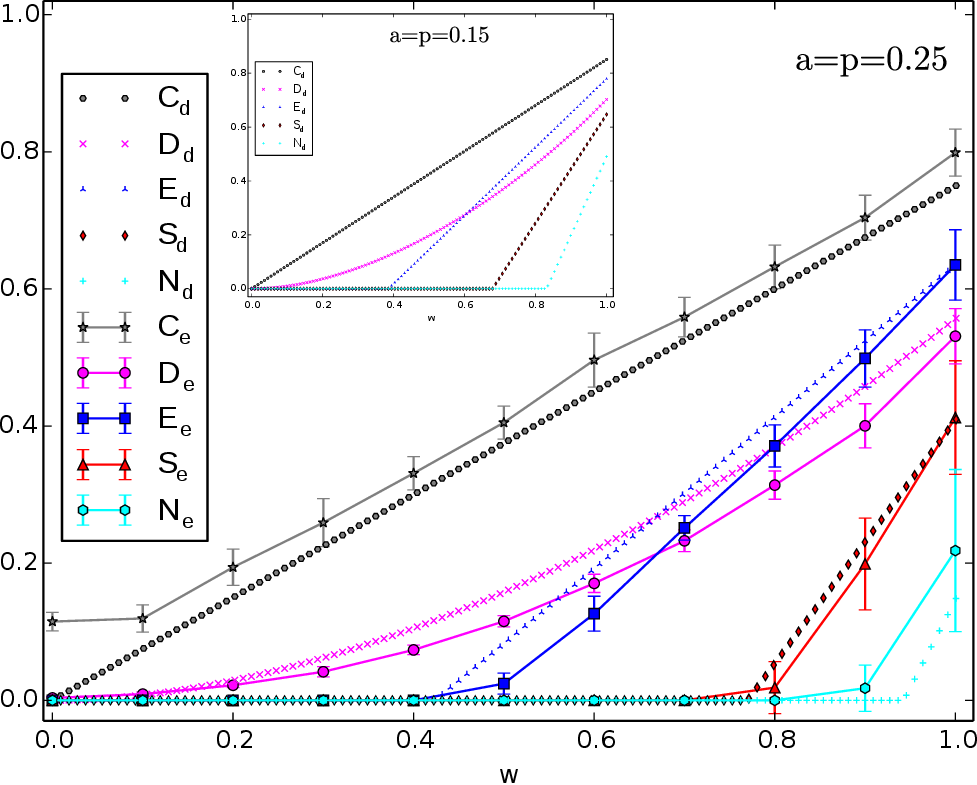}

\caption{(color online) Quantum non-local coherence and quantum correlations
of Werner states with one of the qubits evolved under the composition
of phase damping and amplitude damping channels. Here we used the
same labeling of Fig. \ref{qcorr}, with the subscripts $d$ and $e$
standing for the function calculated using the decohered states of
Eq. (\ref{eq:decoh}) and the experimental-reconstructed states, respectively.
The main figure is for $a=p=0.25$ and in the inset are shown the
corresponding curves for $a=p=0.15$. We see that, for $p=a=0.25$,
although non-local coherence and correlations are under-estimated
while discord, entanglement, and steering are over-estimated by the
theoretical noise model, this model reproduces fairly well the experimentally
observed greater susceptibility of stronger correlations to environmental
interactions.}

\label{decoh}
\end{figure}

It is interesting noticing that not only are the different quantum
resources affected unevenly by those external influences, but the
stronger the quantum correlation is, the more it is impacted. This
fact can be qualitatively well explained in a simplified manner through
the application of the composition of the amplitude damping and phase
damping channels \cite{nilsen,soares-pinto} to one of the qubits
of the theoretical-target Werner states:
\begin{equation}
\rho_{w}^{d}(a,p)=\sum_{j=0}^{2}K_{j}(a,p)\otimes\sigma_{0}\rho_{w}K_{j}^{\dagger}(a,p)\otimes\sigma_{0},\label{eq:decoh}
\end{equation}
with the Kraus' operators given by: 
\begin{eqnarray}
K_{0}(a,p) & = & \sqrt{p(1-a)}|1\rangle\langle1|,\\
K_{1}(a,p) & = & \sqrt{a}|0\rangle\langle1|,\\
K_{2}(a,p) & = & |0\rangle\langle0|+\sqrt{(1-p)(1-a)}|1\rangle\langle1|.
\end{eqnarray}
The quantum non-local coherence and correlations of $\rho_{w}^{d}(0.25,0.25)$
and of $\rho_{w}^{d}(0.15,0.15)$ are shown in Fig. \ref{decoh}.

The results in Fig. \ref{decoh} show that our simplified model of
Eq. (\ref{eq:decoh}) describes well the main qualitative features
of the experimental data. Besides, we see in the inset of Fig. \ref{decoh}
that by lowering the values of the amplitude and phase noise rates,
one could significantly increase the values of the quantumness functions
of the generated states.

In conclusion, in this article we gave a quantum circuit that can
be used to prepare tunable Bell-diagonal states with a quantum computer.
We implemented this quantum circuit using the IBM Q 5 Yorktown quantum
computer and measured the non-local quantum coherence and discord,
entanglement, steering, and non-local quantum correlations of experimentally
reconstructed Werner states. Even though the noise and imperfections
of the hardware utilized had a quite strong detrimental effect on
the measured quantum correlations, we succeeded in verifying a hierarchy
relation for quantum resources (see e.g. Ref. \cite{costa}) of the
produced states: $N\Rightarrow S\Rightarrow E\Rightarrow D.$ The
simple zero-temperature composite noise model we made to explain the
obtained experimental results indicates that access to quantum computers
with lower noise rates will allow for the application of our quantum
circuit to produce even the strongest kinds of quantum correlation
and also to test several interesting theoretical results that have
been reported in the recent quantum information science literature
using Bell-diagonal states.
\begin{acknowledgments}
This work was supported by the Brazilian National Institute for the Science and Technology of Quantum Information (INCT-IQ), process 465469/2014-0, and by the Coordination for the Improvement of Higher Education Personnel (CAPES).
\end{acknowledgments}


\end{document}